\documentclass[12pt]{article}
\pdfoutput=1
\usepackage{bm,amsmath,amssymb,graphicx}

\begin{document}
\pagenumbering{arabic}
\begin{titlepage}

\title{Early universe nucleosynthesis in massive conformal gravity}

\author{F. F. Faria$\,^{*}$ \\
Centro de Ci\^encias da Natureza, \\
Universidade Estadual do Piau\'i, \\ 
64002-150 Teresina, PI, Brazil}

\date{}
\maketitle

\begin{abstract}
We study the dynamics of the early universe in massive conformal 
gravity. In particular, we show that the theory is consistent with 
the observed values of the primordial abundances of light elements if 
we consider the existence of right-handed sterile neutrinos.
\end{abstract}

\thispagestyle{empty}
\vfill
 \par
\bigskip
\noindent * felfrafar@hotmail.com \par
\end{titlepage}
\newpage


\section{Introduction}
\label{sec1}


It is well known that the standard $\Lambda$CDM cosmological model is 
consistent with most observations of the universe at both early and late times 
\cite{Ries,Agha}. However, for this consistency to occur, a very small value 
for the cosmological constant ($\Lambda$) is required, which by far does not 
match with the huge value predicted by quantum field theory (see \cite{Rugh} 
for a nice review). This discrepancy between the cosmological and quantum 
values of $\Lambda$ is known as the cosmological constant problem \cite{Wein}.
Another important problem of $\Lambda$CDM is that the primordial lithium 
abundance from the early universe nucleosynthesis predicted by it differs by 
about a factor of three from the observed abundance \cite{Cybu}, which is 
known as the lithium problem. Despite several attempts over the years, no 
alternative cosmological model has succeeded in solving these two  
problems and being consistent with other cosmological observations 
at the same time.

One of such models comes from massive conformal 
gravity (MCG), which is a conformally invariant theory of gravity in which 
the gravitational action is the sum of the Weyl action with the 
Einstein-Hilbert action conformally coupled to a scalar field \cite{Faria1}.
Among so many cosmological models, we chose the MCG model because it fits 
well with the Type Ia supernovae (SNIa) data without the cosmological constant 
problem \cite{Faria2}. In addition, the theory is free of the van 
Dam-Veltman-Zakharov (vDVZ) discontinuity \cite{Faria3}, can reproduce the 
orbit of binaries by the emission of gravitational waves \cite{Faria4} and 
is consistent with solar system observations \cite{Faria5}. Furthermore, MCG 
is a power-counting renormalizable \cite{Faria6,Faria7} and unitary 
\cite{Faria8} quantum theory of gravity. 

In this paper, we want to see if the MCG cosmology is consistent with the observed 
primordial abundances of light elements without the lithium problem. In Sec. 2, 
we describe the MCG cosmological equations. In Sec. 3, we derive the matter 
energy-momentum tensor used in the theory. In Sec. 4, we study the dynamics of 
the early MCG universe. In Sec. 5, we compare the early universe nucleosynthesis
of MCG with cosmological observations. In Sec. 6, we analyze the evolution of 
the baryon density of the MCG universe. Finally, in Sec. 7, we present our 
conclusions.


\section{Massive conformal gravity}
\label{sec2}


The total MCG action is given by\footnote{This action is obtained from the 
action of Ref. \cite{Faria3} by rescaling $\varphi \rightarrow 
\left(\sqrt{32\pi G/3}\right)\varphi$ and considering 
 $m = \sqrt{3/64\pi G\alpha}$.} \cite{Faria3}
\begin{equation}
S = \int{d^{4}x} \, \sqrt{-g}\bigg[ \varphi^{2}R 
+ 6 \partial^{\mu}\varphi\partial_{\mu}\varphi 
- \frac{1}{2\alpha^2} C^{\alpha\beta\mu\nu}C_{\alpha\beta\mu\nu} \bigg] 
+ \frac{1}{c}\int{d^{4}x\mathcal{L}_{m}},
\label{1}
\end{equation}
where $\varphi$ is a scalar field called 
dilaton, $\alpha$ is a coupling constant, 
\begin{equation}
C^{\alpha\beta\mu\nu}C_{\alpha\beta\mu\nu} = R^{\alpha\beta\mu\nu}
R_{\alpha\beta\mu\nu} - 4R^{\mu\nu}R_{\mu\nu} + R^2 
+ 2\left(R^{\mu\nu}R_{\mu\nu} - \frac{1}{3}R^{2}\right)
\label{2}
\end{equation}
is the Weyl tensor squared,
$R^{\alpha}\,\!\!_{\mu\beta\nu} 
= \partial_{\beta}\Gamma^{\alpha}_{\mu\nu} + \cdots$ is the Riemann tensor, 
$R_{\mu\nu} = R^{\alpha}\,\!\!_{\mu\alpha\nu}$ is the Ricci tensor, 
$R = g^{\mu\nu}R_{\mu\nu}$ is the scalar curvature, and 
$\mathcal{L}_{m} = \mathcal{L}_{m}(g_{\mu\nu},\Psi)$ is the Lagrangian 
density of the matter field $\Psi$. It is worth noting that besides being 
invariant under coordinate transformations, the action (\ref{1}) is also 
invariant under the conformal transformations
\begin{equation}
\tilde{\Phi} = \Omega(x)^{-\Delta_{\Phi}}\Phi,
\label{3}
\end{equation}
where $\Omega(x)$  is an arbitrary function of the spacetime coordinates, 
and $\Delta_{\Phi}$ is the scaling dimension of the field $\Phi$, whose 
values are $-2$ for the metric field, $0$ for gauge bosons, $1$ for 
scalar fields, and $3/2$ for fermions. 

The variation of (\ref{1}) with respect to $g^{\mu\nu}$ and $\varphi$ gives 
the MCG field equations
\begin{equation}
\varphi^{2}G_{\mu\nu} +  6 \partial_{\mu}\varphi\partial_{\nu}\varphi 
- 3g_{\mu\nu}\partial^{\rho}\varphi\partial_{\rho}\varphi + g_{\mu\nu} 
\nabla^{\rho}\nabla_{\rho} \varphi^{2} 
- \nabla_{\mu}\nabla_{\nu} \varphi^{2}  - \alpha^{-2} W_{\mu\nu} 
= \frac{1}{2c}T_{\mu\nu},
\label{4}
\end{equation}
\begin{equation}
\left(\nabla^{\mu}\nabla_{\mu} - \frac{1}{6}R \right) \varphi = 0,
\label{5}
\end{equation}
where
\begin{eqnarray}
W_{\mu\nu} &=& \nabla^{\rho}\nabla_{\rho}R_{\mu\nu} 
- \frac{1}{3}\nabla_{\mu}\nabla_{\nu}R  -\frac{1}{6}g_{\mu\nu}\nabla^{\rho}
\nabla_{\rho}R + 2R^{\rho\sigma}R_{\mu\rho\nu\sigma} 
-\frac{1}{2}g_{\mu\nu}R^{\rho\sigma}R_{\rho\sigma}  \nonumber \\ &&
- \frac{2}{3}RR_{\mu\nu}  + \frac{1}{6}g_{\mu\nu}R^2
\label{6}
\end{eqnarray}
is the Bach tensor,
\begin{equation}
G_{\mu\nu} = R_{\mu\nu} - \frac{1}{2}g_{\mu\nu}R
\label{7}
\end{equation}
is the Einstein tensor,
\begin{equation}
\nabla^{\rho}\nabla_{\rho} \varphi = 
\frac{1}{\sqrt{-g}}\partial^{\rho}\left( \sqrt{-g} \partial_{\rho}
\varphi \right)
\label{8}
\end{equation} 
is the generally covariant d'Alembertian for a scalar field, and
\begin{equation}
T_{\mu\nu} = - \frac{2}{\sqrt{-g}} \frac{\delta \mathcal{L}_{m}}
{\delta g^{\mu\nu}}
\label{9}
\end{equation}
is the matter energy-momentum tensor.

Before we proceed, it is important to note that both the symmetries of the 
theory allow us to introduce in (\ref{1}) a quartic self-interacting term of 
the dilaton $\lambda\int{\sqrt{-g} \varphi^4}$ as well as interaction terms 
of the dilaton with the matter fields. In the case of the dilaton 
self-interaction term, we do not include it in the MCG action because this 
inclusion makes the flat metric no longer a solution of the field equations, 
which invalidates the S-matrix formulation. Although such a term is 
reintroduced in the effective action by quantum corrections, we can consider 
the renormalized value of the coupling constant $\lambda$ equal zero so that 
the self-interacting term is present in the renormalized action only to cancel 
out the corresponding divergent term. In addition, we neglect the couplings between 
the dilaton and the matter fields because they make the field equation (\ref{5}) 
no longer valid. This equation is fundamental to cancel non-renormalizable 
divergent terms that appear in the effective action \cite{Faria9}.

At scales below the Planck scale, the dilaton field acquires a spontaneously 
broken constant vacuum expectation value $\varphi_{0}$ \cite{Matsuo}. In 
this case, the field equations (\ref{4}) and (\ref{5}) become 
\begin{equation}
\varphi_{0}^{2}G_{\mu\nu} - \alpha^{-2} W_{\mu\nu} = \frac{1}{2c}T_{\mu\nu},
\label{10}
\end{equation}
\begin{equation}
R = 0.
\label{11}
\end{equation} 
In addition, for $\varphi = \varphi_{0}$, the MCG line element 
$ds^2 = \left(\varphi/\varphi_{0}\right)^{2}g_{\mu\nu}dx^{\mu}dx^{\nu}$ 
reduces to  
\begin{equation}
ds^2 = g_{\mu\nu}dx^{\mu}dx^{\nu}.
\label{12}
\end{equation}
The full dynamics of the MCG universe can be described by
(\ref{10})-(\ref{12}) without loss of generality.


\section{Dynamical perfect fluid}
\label{sec3}


In order to find the MCG matter energy-momentum tensor, we 
consider the conformally invariant matter Lagrangian density \cite{Man1}
\begin{equation}
\mathcal{L}_{m} = - \sqrt{-g}c\bigg[S^{2}R + 6\partial^{\mu}S\partial_{\mu}S 
+ \lambda S^{4} + \frac{i}{2}\hbar\left(\, \overline{\psi}
\gamma^{\mu}D_{\mu}\psi - D_{\mu}\overline{\psi}\gamma^{\mu}\psi \right) 
- \hbar\mu S\overline{\psi}\psi\bigg],
\label{13}
\end{equation}
where $S$ is a scalar Higgs field\footnote{Although 
the Higgs field is actually a doublet, and it is more likely that we must have 
two more scalar fields to get the correct quantum phenomenology at low energies 
\cite{Hel}, considering only a scalar Higgs field will not change the 
classical results of the theory.}, $\lambda$ and $\mu$ are coupling constants, 
$\overline{\psi} = \psi^{\dagger}
\gamma^{0}$ is the adjoint fermion field, $D_{\mu} = \partial_{\mu} 
+ [\gamma^{\nu},\partial_{\mu}\gamma_{\nu}]/8 - [\gamma^{\nu},\gamma_{\lambda}]
\Gamma^{\lambda}\,\!\!_{\mu\nu}/8$ ($\Gamma^{\lambda}\,\!\!_{\mu\nu}$ is the 
Levi-Civita connection), and $\gamma^{\mu}$ 
are the general relativistic Dirac matrices, which satisfy the anti-commutation 
relation $\{\gamma^{\mu},\gamma^{\nu}\} = 2g^{\mu\nu}$. 

By varying (\ref{13}) with respect to $S$, $\overline{\psi}$ 
and $\psi$, we obtain the field equations
\begin{equation}
12\nabla^{\mu}\nabla_{\mu}S - 2RS
- 4\lambda S^3 + \hbar\mu\overline{\psi}\psi = 0,
\label{14}
\end{equation}
\begin{equation}
i\gamma^{\mu}D_{\mu}\psi - \mu S \psi = 0,
\label{15}
\end{equation}
\begin{equation}
iD_{\mu}\overline{\psi}\gamma^{\mu} + \mu S \overline{\psi} = 0.
\label{16}
\end{equation}
Additionally, the substitution of (\ref{13}) into (\ref{9}) gives
\begin{eqnarray}
\frac{T_{\mu\nu}}{c} &=& 12\partial_{\mu}S\partial_{\nu}S 
- 6g_{\mu\nu}\partial^{\rho}S\partial_{\rho}S + 2 g_{\mu\nu} 
\nabla^{\rho}\nabla_{\rho} S^{2} 
- 2 \nabla_{\mu}\nabla_{\nu} S^{2} \nonumber \\ &&
+ \, 2S^{2}G_{\mu\nu}
- g_{\mu\nu}\left[\lambda S^4 + \frac{i}{2}\hbar\left(\, \overline{\psi}
\gamma^{\rho}D_{\rho}\psi - D_{\rho}\overline{\psi}\gamma^{\rho}\psi \right) 
- \hbar\mu S\overline{\psi}\psi\right] \nonumber \\ &&
+ \, \frac{i}{4}\hbar\big(\, \overline{\psi}
\gamma_{\mu}D_{\nu}\psi - D_{\nu}\overline{\psi}\gamma_{\mu}\psi 
+ \overline{\psi}\gamma_{\nu}D_{\mu}\psi - D_{\mu}\overline{\psi}\gamma_{\nu}
\psi \big).
\label{17}
\end{eqnarray}
Then, using (\ref{14})-(\ref{16}) and $\nabla_{\mu}\nabla_{\nu}S^2 = 
2(S\nabla_{\mu}\nabla_{\nu}S + \partial_{\mu}S\partial_{\nu}S)$ in (\ref{17}), 
we find the energy-momentum tensor
\begin{eqnarray}
T_{\mu\nu} &=& c \left(8\partial_{\mu}S\partial_{\nu}S - 2g_{\mu\nu}\partial^{\rho}S
\partial_{\rho}S- 4 S\nabla_{\mu}\nabla_{\nu} S  
+ g_{\mu\nu}S\nabla^{\rho}\nabla_{\rho} S\right) \nonumber \\ &&
+ \, 2cS^{2}\left(R_{\mu\nu} - \frac{1}{4}g_{\mu\nu}R\right) 
+ T^{f}_{\mu\nu},
\label{18}
\end{eqnarray}
where
\begin{equation}
T^{f}_{\mu\nu} = \frac{i}{4}c\hbar\big(\, \overline{\psi}
\gamma_{\mu}D_{\nu}\psi - D_{\nu}\overline{\psi}\gamma_{\mu}\psi 
+ \overline{\psi}\gamma_{\nu}D_{\mu}\psi - D_{\mu}\overline{\psi}\gamma_{\nu}
\psi \big) - \frac{1}{4} g_{\mu\nu}c\hbar\mu S\overline{\psi}\psi
\label{19}
\end{equation}
is the fermion energy-momentum tensor.

Considering that, at scales below the electroweak scale, the Higgs field 
acquires a spontaneously broken constant vacuum expectation value $S_{0}$, and 
making some algebra, we find that (\ref{15}) and (\ref{18}) become
\begin{equation}
\left[D^{\mu}D_{\mu} - \left(\frac{mc}{\hbar}\right)^2\right] \psi = 0,
\label{20}
\end{equation}
\begin{equation}
T_{\mu\nu}(S_{0},g_{\mu\nu}) = 2cS_{0}^{2}\left(R_{\mu\nu} 
- \frac{1}{4}g_{\mu\nu}R\right) + T^{f}_{\mu\nu}(S_{0},g_{\mu\nu}),
\label{21}
\end{equation}
where
\begin{equation}
T^{f}_{\mu\nu}(S_{0},g_{\mu\nu}) = \frac{i}{4}c\hbar\big(\, \overline{\psi}
\gamma_{\mu}D_{\nu}\psi - D_{\nu}\overline{\psi}\gamma_{\mu}\psi 
+ \overline{\psi}\gamma_{\nu}D_{\mu}\psi - D_{\mu}\overline{\psi}\gamma_{\nu}
\psi \big) - \frac{1}{4}g_{\mu\nu}m c^2\overline{\psi}\psi,
\label{22}
\end{equation}
with $m = \mu S_{0} \hbar/c$ being the fermion mass. In flat spacetime, is not 
difficult to see that (\ref{20}) and (\ref{22}) reduce to
\begin{equation}
\left[\partial^{\mu}\partial_{\mu} - \left(\frac{mc}{\hbar}\right)^2\right] \psi = 0,
\label{23}
\end{equation}
\begin{equation}
T^{f}_{\mu\nu}(S_{0},\eta_{\mu\nu}) = \frac{i}{4}c\hbar\big(\, \overline{\psi}
\gamma_{\mu}\partial_{\nu}\psi - \partial_{\nu}\overline{\psi}\gamma_{\mu}\psi 
+ \overline{\psi}\gamma_{\nu}\partial_{\mu}\psi - \partial_{\mu}\overline{\psi}
\gamma_{\nu}\psi \big) - \frac{1}{4}\eta_{\mu\nu}m c^2\overline{\psi}\psi,
\label{24}
\end{equation}
where now the  Dirac matrices satisfy the anti-commutation 
relation $\{\gamma^{\mu},\gamma^{\nu}\} = 2\eta^{\mu\nu}$. 

The normalized plane wave solution to (\ref{23}) is given by
\begin{equation}
\psi = \frac{1}{\sqrt{VE_{k}}}\, u_{k} \, e^{ik_{\mu}x^\mu},
\label{25}
\end{equation}
where $V$ is the volume, $E_{k} = \sqrt{k^2c^2 + m^2c^4}$ is the energy, 
$u_{k}$ is a spinor which satisfies 
$\left[\gamma^{\mu}k_{\mu} + mc/\hbar \right] u_{k} = 0$, 
and $k_{\mu} = (E_{k}/c\hbar, \vec{k}/\hbar)$ is the wave 
vector, with $\vec{k}$ being the momentum and $k = |\vec{k}|$. By substituting 
(\ref{25}) and its adjoint into (\ref{24}), and using $\overline{u}_{k}u_{k} 
= - mc^2$, we obtain
\begin{equation}
T^{f}_{\mu\nu}(S_{0},\eta_{\mu\nu}) = \left(\frac{c^2 \hbar^2 }{VE_{k}}\right)
k_{\mu}k_{\nu} + \left(\frac{m^2 c^4}{4VE_{k}}\right)\eta_{\mu\nu}.
\label{26}
\end{equation}
Incoherently adding to (\ref{26}) the individual contributions of a set of six 
plane waves moving in the $\pm \, x$, $\pm \, y$ and $\pm \, z$ directions, all with 
the same $E_k$ and $k$, we can write the energy-momentum tensor (\ref{26}) in the 
perfect fluid form
\begin{equation}
T^{f}_{\mu\nu}(S_{0},\eta_{\mu\nu}) = \left( \rho + \frac{p}{c^2} \right)
u_{\mu}u_{\nu} + \eta_{\mu\nu}p + \eta_{\mu\nu}c^2\rho_{\Lambda},
\label{27}
\end{equation}
where 
\begin{equation}
c^2\rho = \frac{6E_k}{V}
\label{28}
\end{equation}
is the energy density of the fluid,
\begin{equation}
p = \frac{2k^2c^2}{VE_k}
\label{29}
\end{equation}
is the pressure of the fluid,
\begin{equation}
c^2\rho_{\Lambda} = \frac{3m^2c^4}{2V E_k}
\label{30}
\end{equation}
is the vacuum energy (dark energy) density, and $u^{\mu}$ is the
four-velocity of the fluid, which is normalized to $u^{\mu}u_{\mu} = - c^2$.
It follows from (\ref{28})-(\ref{30}) that
\begin{equation}
p = 0, \ \ \ \ \ \ \ \rho_{\Lambda} = \frac{1}{4}\rho,
\label{31}
\end{equation}
for a non-relativistic perfect fluid ($k^2c^2 \ll m^2c^4$), and
\begin{equation}
p = \frac{1}{3}c^2\rho, \ \ \ \ \ \ \ \rho_{\Lambda} = 0,
\label{32}
\end{equation}
for a relativistic perfect fluid ($k^2c^2 \gg m^2c^4$). 

In curved spacetime, the perfect fluid energy-momentum tensor (\ref{27}) 
is generalized to
\begin{equation}
T^{f}_{\mu\nu}(S_{0},g_{\mu\nu}) = \left( \rho + \frac{p}{c^2} \right)
u_{\mu}u_{\nu} + g_{\mu\nu}p + g_{\mu\nu}c^2\rho_{\Lambda}.
\label{33}
\end{equation}
Finally, the insertion of (\ref{33}) into (\ref{21}) gives the 
energy-momentum tensor of a dynamical perfect fluid
\begin{equation}
T_{\mu\nu}(S_{0},g_{\mu\nu}) = 2cS_{0}^{2}\left(R_{\mu\nu} 
- \frac{1}{4}g_{\mu\nu}R\right) + \left( \rho + \frac{p}{c^2} \right)
u_{\mu}u_{\nu} + g_{\mu\nu}p + g_{\mu\nu}c^2\rho_{\Lambda}.
\label{34}
\end{equation}
Taking the trace of (\ref{34}), and substituting into the trace of (\ref{10}), 
whose left hand side is zero due to the field equation (\ref{11}) and the 
tracelessness of the Bach tensor ($W = g^{\mu\nu}W_{\mu\nu} =0$), we arrive at
\begin{equation}
T = g^{\mu\nu}T_{\mu\nu} = 3p - c^2\rho + 4 c^2\rho_{\Lambda} = 0.
\label{35}
\end{equation}
We can see from (\ref{31}) and (\ref{32}) that both non-relativistic 
and relativistic perfect fluids satisfies the  tracelessness relation 
(\ref{35}). For simplicity, we could isolate $\rho_{\Lambda}$ in (\ref{35}) 
and replace it in (\ref{34}) as done in Ref. \cite{Faria2}. In this case, it is 
made clear that the vacuum energy density does not contribute directly to the dynamic 
evolution of the MCG universe, which solves the cosmological constant problem found in 
the $\Lambda$CDM model. However, here we will keep $\rho_{\Lambda}$ so we don't 
miss any physical details during the calculations.

By substituting (\ref{34}) into (\ref{10}), and considering (\ref{11}), we find
\begin{equation}
\left(\varphi_{0}^{2}-S_{0}^{2}\right)R_{\mu\nu} 
- \alpha^{-2}W_{\mu\nu} = \frac{1}{2c}\left[ \left( \rho 
+ \frac{p}{c^2} \right)u_{\mu}u_{\nu} + g_{\mu\nu}p 
+ g_{\mu\nu}c^2\rho_{\Lambda}\right],
\label{36}
\end{equation}
which is the field equation that we will use in the study of the 
dynamics of the early MCG universe in the next section. But before that, it is 
important to compare MCG with another conformally invariant theory 
of gravity called conformal gravity (CG)\footnote{Although the difference 
between the two theories is quite obvious, as we will readily show next, 
MCG is often confused with CG. Perhaps this is because CG is much older 
and known than MCG.}, whose action is given by \cite{Man2}
\begin{equation}
S = - \frac{1}{2\alpha^2}\int{d^{4}x} \, \sqrt{-g} 
\left(C^{\alpha\beta\mu\nu}C_{\alpha\beta\mu\nu} \right)
+ \frac{1}{c}\int{d^{4}x\mathcal{L}_{m}}.
\label{37}
\end{equation}
By varying (\ref{37}) with respect to $g_{\mu\nu}$, we obtain the field equation
\begin{equation}
 - \alpha^{-2}W_{\mu\nu} = \frac{1}{2c}T_{\mu\nu},
\label{38}
\end{equation}
where $T_{\mu\nu}$ is given by (\ref{18}). We can easily see the difference 
between the two theories by comparing (\ref{38}) with (\ref{10}) 
and (\ref{11}). Just to stay within the scope of this paper, it is worth noting 
that CG does not pass the early universe nucleosynthesis test \cite{Knox}.


\section{Early universe}
\label{sec4}


As usual, we consider that the geometry of the universe is described by the 
Friedmann–Lema\^itre–Robertson–Walker (FLRW) line element
\begin{equation}
ds^{2} = - c^2dt^{2} + a(t)^2\left( \frac{dr^{2}}{1-Kr^{2}} +r^{2}d\theta^{2} 
+ r^{2}\sin^{2}\theta d\phi^{2} \right),
\label{39}
\end{equation}
where $a = a(t)$ is the scale factor and $K=$ -1, 0 or 1 is the spatial 
curvature. By substituting (\ref{39}) and the fluid four-velocity 
$u^{\mu} = (c, 0, 0 ,0)$ into (\ref{36}), we obtain\footnote{It is worth 
noting that $W_{\mu\nu} = 0$ for the FLRW spacetime.}
\begin{equation}
\frac{\ddot{a}}{a} = -\frac{c}{6\left(\varphi_{0}^{2}-S_{0}^{2}\right)}
\left(c^2\rho - c^2\rho_{\Lambda} \right),
\label{40}
\end{equation}
\begin{equation}
\frac{\ddot{a}}{a} + 2\left( \frac{\dot{a}}{a}\right)^2 + 2 \frac{Kc^2}{a^2}
= \frac{c}{2\left(\varphi_{0}^{2}-S_{0}^{2}\right)}
\left(p + c^{2}\rho_{\Lambda} \right),
\label{41}
\end{equation}
where the dot denotes $d/dt$. 

Subtracting (\ref{40}) from (\ref{41}), and considering that\footnote{This 
value of $\varphi_{0}$ is necessary for the theory to be consistent with solar 
system observations \cite{Faria5}.}
\begin{equation}
\varphi_{0}^{2} = \frac{3c^3}{32\pi G} \gg S_{0}^{2},
\label{42}
\end{equation} 
we obtain 
\begin{equation}
\left(\frac{\dot{a}}{a}\right)^2   
=  \frac{8 \pi G}{9c^2}\left(c^{2}\rho + 3p +  2c^{2}\rho_{\Lambda} \right) 
- \frac{Kc^2}{a^2}.
\label{43}
\end{equation}
The combination of (\ref{43}) with (\ref{40}) then gives the energy continuity 
equation 
\begin{equation}
c^2\dot{\rho} + 3\frac{\dot{a}}{a}\left(c^2\rho + p\right) 
- c^2\dot{\rho}_{\Lambda} = 0,
\label{44}
\end{equation}
which can also be obtained by the conservation law $\nabla^\mu{T^{f}_{\mu\nu}} 
= 0$, with $T^{f}_{\mu\nu}$ being the perfect fluid 
energy-momentum tensor (\ref{33}). 

Using either (\ref{31}) or (\ref{32}) in (\ref{44}), we get
\begin{equation}
\dot{\rho} + 4\frac{\dot{a}}{a}\rho = 0,
\label{45}
\end{equation}
which, consequently, is valid for both non-relativistic and relativistic 
dynamical perfect fluids. As usual, we can write the solution to (\ref{45}) 
in the form
\begin{equation}
\rho = \rho_{0}\left(\frac{a_{0}}{a}\right)^{4},
\label{46}
\end{equation}
where, from now on, the subscript $0$ denotes values at the present time 
$t_{0}$.
 
In the case of the early universe, which is composed by a very hot plasma 
dominated by relativistic particles (radiation), we find that (\ref{43}) 
becomes 
\begin{equation}
\dot{a}^2   
=  \frac{16 \pi Ga_{0}^4}{9a^2}\rho_{r0}- Kc^2.
\label{47}
\end{equation}
where we used (\ref{32}) and (\ref{46}), with $\rho_{r}$ being the mass 
density of the radiation. Since $a$ is small in the early universe, we can 
neglect the curvature term on the right hand side of (\ref{47}) and write it 
in the approximate form
\begin{equation}
\dot{a}^2   =  \frac{16 \pi Ga_{0}^4}{9a^2}\rho_{r0},
\label{48}
\end{equation}
whose solution is given by
\begin{equation}
a(t)  = \left(\frac{64 \pi Ga_{0}^4\rho_{r0}}{9}\right)^{1/4}t^{1/2}.
\label{49}
\end{equation}

Finally, inserting (\ref{49}) into the Hubble constant
\begin{equation}
H  = \frac{\dot{a}}{a},
\label{50}
\end{equation}
we obtain
\begin{equation}
H  = \frac{1}{2t},
\label{51}
\end{equation}
which is the same relation between the Hubble constant and time that occurs in 
the early $\Lambda$CDM universe. However, since the MCG scale factor (\ref{49}) 
is equal $0.9$ times the value of the $\Lambda$CDM scale factor, the expansion of 
the early MCG universe is slower than the expansion of the early $\Lambda$CDM 
universe, which will give a difference in the values of the two Hubble constants, 
as we will show in the next section. 


\section{Nucleosynthesis}
\label{sec5}


The abundances of light chemical elements in the early universe are mainly 
determined by one cosmological parameter, namely, the baryon-to-photon ratio 
$\eta = n_{b}/n_{\gamma}$, where $n_{b}$ and $n_{\gamma}$ are the number 
densities of baryons and photons in the universe. 
As usual, to find $\eta$ we must first write the Hubble constant in function 
of temperature $T$ using the Stefan-Boltzmann law
\begin{equation}
\rho_{r} = \left(\frac{g_{*}a_{\mathcal{B}}}{2c^2}\right)T^4,
\label{52}
\end{equation}
where $a_{\mathcal{B}}$ is the radiation energy constant and $g_{*}$ counts the 
number of relativistic particle species determining the energy density in 
radiation. Substituting (\ref{52}) and (\ref{49}) into (\ref{46}), we obtain
\begin{equation}
t = \left(\frac{9c^2}{32 \pi G g_{*}a_{\mathcal{B}}}\right)^{1/2}\frac{1}{T^{2}}.
\label{53}
\end{equation}
It then follows from (\ref{51}) and (\ref{53}) that
\begin{equation}
H = \left(\frac{8 \pi G g_* a_{\mathcal{B}}}{9c^2}\right)^{1/2}T^{2},
\label{54}
\end{equation}
which is equal $0.82$ times the value of the $\Lambda$CDM Hubble constant.

In order to describe the thermal history of the early MCG universe, we must
compare the Hubble constant in the form (\ref{54}) with the collision rate of 
particle interactions 
\begin{equation}
\Gamma = n\sigma v,
\label{55}
\end{equation}
where $n$ is the number density of particles, $\sigma$ is their interaction 
cross section and $v$ is the average velocity of the particles. A specific
temperature that is of particular importance for the outcome of the early 
universe nucleosynthesis (EUN) is the one at which the thermal equilibrium 
between neutrons and protons begins to break down, which happens when 
$H \sim \Gamma_{\nu}$, where 
\begin{equation}
\Gamma_{\nu} \approx \frac{G_{F}^2}{c^6 \hbar^7}(k_{\mathcal{B}}T)^5
\label{56}
\end{equation}
is the collision rate of a neutrino with electrons or positrons, with 
$G_{F}$ being the Fermi constant and $k_{\mathcal{B}}$ the Boltzmann 
constant. 

By equating (\ref{54}) with (\ref{56}), and assuming that at the onset 
of the electron-positron annihilation the remaining relativistic particles 
are photons, electrons, positrons and left-handed neutrinos, for which 
$g_* =  10.75$, we obtain
\begin{equation}
k_{\mathcal{B}}T_{\mathrm{eq}} = 0.75 \, \mathrm{MeV}.
\label{57}
\end{equation}
We can see from (\ref{57}) that the thermal equilibrium between neutrons 
and protons is maintained at temperatures above $T_{\mathrm{eq}} =  8.7 
\times 10^{9} \, \mathrm{K}$ in the early MCG universe. At that time, 
the neutron-to-proton ratio was 
\begin{equation}
\left(\frac{n_{n}}{n_{p}}\right)_{\mathrm{eq}} 
= e^{-Q/k_{\mathcal{B}}T_{\mathrm{eq}}} = 0.178,
\label{58}
\end{equation}
where we used (\ref{57}) and the neutron-proton energy difference 
$Q = 1.239 \, \mathrm{MeV}$. Using (\ref{58}), 
we can make a rough estimate that the final freeze-out neutron 
abundance is given by
\begin{equation}
X^{\infty}_{n} \sim X^{\mathrm{eq}}_{n} 
= \frac{ e^{-Q/k_{\mathcal{B}}T_{\mathrm{eq}}}}{1 
+ e^{-Q/k_{\mathcal{B}}T_{\mathrm{eq}}}} = 0.15.
\label{59}
\end{equation}
Including the neutron decay in our calculation, we find
\begin{equation}
X_{n}(t) = X^{\infty}_{n}e^{-t/\tau_{n}} = 0.15 \, e^{-t/\tau_{n}},
\label{60}
\end{equation}
where $\tau_{n} = 879.4 \, \mathrm{s}$ is the neutron mean lifetime \cite{Tana}.

The first light element formed in the early universe was deuterium (D), 
whose ratio to proton is approximately  given by
\begin{equation}
\frac{n_{D}}{n_{p}} \approx 6.9 \eta \left( \frac{k_{\mathcal{B}}T}{m_{n}c^2} 
\right)^{3/2}\mathrm{exp}\left(\frac{B_{D}}{k_{\mathcal{B}}T}\right),
\label{61}
\end{equation}
where we used (\ref{58}) and $B_{D} = 2.2 \, \mathrm{MeV} $ is the binding 
energy of deuterium. Noting that the EUN starts when $n_{D} \sim n_{p}$, it 
follows from (\ref{61}) that
\begin{equation}
6.9 \eta_\mathrm{EUN} \left( \frac{k_{\mathcal{B}}T_\mathrm{EUN}}{m_{n}c^2} 
\right)^{3/2}\mathrm{exp}\left(\frac{B_{D}}{k_{\mathcal{B}}T_\mathrm{EUN}}
\right) \approx 1,
\label{62}
\end{equation}
where $\eta_\mathrm{EUN}$ and $T_\mathrm{EUN}$ are the baryon-to-photon ratio 
and temperature of the EUN. We can 
see from (\ref{62}) that we need the value of $T_\mathrm{EUN}$ to find 
$\eta_\mathrm{EUN}$. Fortunately, we can find such value from the 
primordial helium ($\mathrm{^4He}$) abundance
\begin{equation}
Y_{P} \equiv \frac{4n_{\mathrm{^4\!He}}}{n_\mathrm{H}} 
= \frac{2 X_{n}(t_\mathrm{EUN})}{1-X_{n}(t_\mathrm{EUN})},
\label{63}
\end{equation}
where $t_\mathrm{EUN}$ is the time of the EUN. 

The substitution of (\ref{60}) and the observed value of the helium 
abundance $Y_P = 0.245$ \cite{Aver} into (\ref{63}) gives 
\begin{equation}
t_\mathrm{EUN} \approx 279.7 \, \mathrm{s}.
\label{64}
\end{equation}
Then, by inserting (\ref{64}) into (\ref{53}), and 
considering that 
the electrons and protons are no longer relativistic after their 
annihilation, which gives $g_* = 3.36$, we obtain
\begin{equation}
T_\mathrm{EUN} \approx 8.8\times 10^8 \, \mathrm{K}.
\label{65}
\end{equation}
Finally, using (\ref{65}) in (\ref{62}), we arrive at
\begin{equation}
\eta_\mathrm{EUN} \approx 5.12 \times 10^{-8},
\label{66}
\end{equation}
which produces abundances of other light elements besides helium orders 
of magnitude below the primordial abundances inferred from current observations 
\cite{Zyla}. However, this result does not automatically rule out MCG. If we 
consider that the theory has low energy ($\lesssim$ eV) right-handed sterile 
neutrinos\footnote{The existence 
of such neutrinos is allowed by the symmetries of the theory and may be 
responsible for the small masses of the left-handed neutrinos found in nature 
\cite{Meissner}.}, 
then we must replace $g_* = 10.75$ by $g_* = 16.125$ prior to the electron-positron 
annihilation and $g_* = 3.36$ by $g_* = 5.04$ after the electron-positron 
annihilation due to the contribution of the sterile neutrinos to the relativistic 
energy content of the universe. These replacements lead to the standard value
\begin{equation}
\eta_\mathrm{EUN} \approx 6 \times 10^{-10},
\label{67}
\end{equation}
which is consistent with the observed abundances of all light elements with the 
exception of lithium\footnote{It is possible that the decay of the sterile 
neutrinos solves the inconsistency between the predicted and observed values 
of the lithium abundance \cite{Salv}.}.


\section{Baryon density}
\label{sec6}


Another important cosmological parameter that is determined by $\eta$ is the 
baryon mass density $\rho_{b}$ of the universe. In order to find the relation 
between these two parameters in the MCG universe, we start from the 
definitions of the baryon and photon number densities
\begin{equation}
n_{b} = \frac{\rho_{b}}{m_N},
\label{68}
\end{equation}
\begin{equation}
n_{\gamma} = 2\zeta(3)\frac{8\pi}{c^3}\left( \frac{k_{\mathcal{B}}T}{h} 
\right)^3 \approx 2\times 10^{7} T^3,
\label{69}
\end{equation}
where $m_N$ is the nucleons mass. The combination of (\ref{68}), (\ref{69}) 
and (\ref{52}), with $g_* = 2$, then gives the relation
\begin{equation}
\eta = \frac{a_{\mathcal{B}}}{2\times 10^7 m_N c^2}
\frac{\rho_{b}}{\rho_{\gamma}} \, T,
\label{70}
\end{equation}
which is valid for any cosmological model. Noting that both $\rho_{b}$ and 
$\rho_{\gamma}$ obey (\ref{46}) in MCG, we can write (\ref{70}) in the form
\begin{equation}
\eta = \frac{a_{\mathcal{B}}}{2\times 10^7 m_N c^2}\frac{\rho_{b0}}
{\rho_{\gamma 0}}\, T,
\label{71}
\end{equation}
which means that the baryon-to-photon ratio evolves over time in the MCG 
universe\footnote{It would be important 
to check if (\ref{71}) at the time of recombination is consistent with 
the value of $\eta$  measured by cosmic microwave background (CMB) 
anisotropies. However, a theory for the growth of inhomogeneities in MCG has 
not yet been developed due to the complexity generated by the contribution 
of the Bach tensor in (\ref{10}). Therefore, we will leave this analysis 
for future works.}, different to what happens in the $\Lambda$CDM universe 
where $\eta$ is constant after the EUN.
  
Using the current temperature of the universe 
$T_0 = 2.73$ K  in (\ref{52}), with $g_* = 2$, we find
\begin{equation}
\rho_{\gamma 0} = 4.65 \times 10^{-31} \,\, \mathrm{kg/m^3}.
\label{72}
\end{equation}
In addition, the use of (\ref{67}) in (\ref{62}), with $6.9$ replaced by $6.5$ 
due to the different value of (\ref{58}) which leads to (\ref{67}), gives
\begin{equation}
T_\mathrm{EUN} \approx 7.56 \times 10^{8} \,\, \mathrm{K}.
\label{73}
\end{equation}
Finally, substituting (\ref{67}), (\ref{72}) and (\ref{73}) into (\ref{71}),
we obtain the current baryon mass density
\begin{equation}
\rho_{b 0} = 1.46 \times 10^{-36} \,\, \mathrm{kg/m^3}.
\label{74}
\end{equation}
Since $\rho_{r}$ and $\rho_{b}$ evolve at the same rate in MCG, it 
follows from (\ref{72}) and (\ref{74}) that radiation always dominates the 
MCG universe. 

In fact, the scale factor is big at late times such that we 
can neglect the density term on the right hand side of (\ref{47}), which 
makes the late MCG universe curvature dominated. In this case,
we must impose $K = -1$, which gives the approximated solution  
\begin{equation}
a(t) = ct
\label{75}
\end{equation} 
in the late MCG universe. It is not difficult to show that for an open universe 
with the scale factor (\ref{75}) such as the late MCG universe, we have the 
luminosity distance
\begin{equation}
d_{L}(z) = \frac{c}{H_{0}}\left[\frac{(1+z)^2 - 1}{2}\right], 
\label{76}
\end{equation}
which fits well to SNIa data\footnote{It is worth noting 
that the density term has not been neglected in Ref. \cite{Faria1}, which in 
practice does not change the SNIa data fitting.} \cite{Faria1}. We intend to 
check if (\ref{75}) provides good fits to other low redshift data in future 
works.

Just to finish, it is important to note that the evolution of 
the baryon-to-photon ratio (\ref{71}) causes the number of baryons $N_{b}$ to 
decrease over time in the MCG universe. We can see this explicitly by 
substituting (\ref{46}) and $V \sim a^3$ in
\begin{equation}
N_{b} = n_{b}V = \frac{\rho_b V}{m_N}, 
\label{77}
\end{equation}
which gives
\begin{equation}
N_{b} \sim \frac{\rho_{b0} a_0^4}{m_N a}.
\label{78}
\end{equation}
Using (\ref{75}), we find that the number of baryons evolves over time 
according to
\begin{equation}
N_{b} \sim \left(\frac{\rho_{b0} c^3 t_0^4}{m_N}\right)t^{-1}
\label{79}
\end{equation}
in the late MCG universe. 

It follows from the energy continuity equation (\ref{44}) that
\begin{equation}
\dot{\rho}_b + 3H\rho_b = \dot{\rho}_{\Lambda}.
\label{80}
\end{equation}
By comparing (\ref{80}) with the standard adiabatic conservation equation, and 
noting that $\dot{\rho}_{\Lambda} < 0$, we conclude that the decrease 
in the number of baryons (\ref{79}) is due to the decay of the baryons into 
dynamic vacuum\footnote{This decaying process can be accounted by the 
Yukawa interaction $\mu S\overline{\psi}\psi$ in (\ref{19}).}, which clearly 
leads to a violation of the conservation of the quantum numbers. However, we 
can see from (\ref{79}) that the variation of the number of baryons should only 
be significant on cosmological time scales, which makes the decay of baryons 
into vacuum not observable in the laboratory. 

On the other hand, the non-conservation of baryons can have an important impact 
on the evolution of inhomogeneous structures of the universe from the end of 
recombination until today. Due to the decrease in the amount of baryons in the MCG 
universe, it is expected that the formation of structures happen much later than is 
observed or not happen at all. However, the evolution of cosmological structures 
does not depend only on baryons but also on dark matter, whose existence is 
necessary in MCG to explain the galaxy rotation curves and the deflection of 
light by galaxies \cite{Faria5}. Therefore, although the theory possibly has 
an extra scalar field that is a good candidate for dark matter \cite{Faria9}, 
much still has to be studied to find out if the evolution of cosmological 
structures predicted by MCG is consistent with observations or not.


\section{Final remarks}
\label{sec7}


Here we have shown that the abundances of light elements, including 
lithium, predicted by the early MCG cosmology are consistent with the 
observed values provided the theory has right-handed sterile 
neutrinos, which is allowed by the symmetries of the theory. Even though we 
still need to check the existence of such neutrinos in experiments like the 
Mini Booster Neutrino Experiment (MiniBooNE) \cite{Agui}, this result is quite 
encouraging for us to continue with the study of the theory.  
   
In addition, it was shown in this paper that the 
baryon-to-photon ratio of the MCG universe evolves over time. Although 
further studies are needed to verify whether this evolution is consistent with 
the value of the baryon-to-photon ratio determined by the CMB anisotropies, who 
knows it solves other early universe problems found in the $\Lambda$CDM model 
such as the baryon asymmetry problem. We intend to study this and other 
MCG cosmological predictions in future works.



\begin{thebibliography}{99}

\bibitem{Ries}
A.G. Riess et al., Astron. J. \textbf{116}, 1009 (1998); S. Perlmutter et al., 
ApJ \textbf{517}, 565 (1999).

\bibitem{Agha}
N. Aghanim et al. [Planck Collab.], Planck 2018 results. VI. Cosmological parameters, 
Astron. Astrophys. \textbf{641}, A6 (2020); Astron. Astrophys. \textbf{652}, C4 (2021).

\bibitem{Rugh}
S. E. Rugh and H. Zinkernagel, Stud. Hist. Phil. Sci. B \textbf{33}, 663 (2002).

\bibitem{Wein}
S. Weinberg, Rev. Mod. Phys. \textbf{61}, 1 (1989).

\bibitem{Cybu}
R. H. Cyburt, B. D. Fields, K. A. Olive and T.-H. Yeh, Rev. Mod. Phys. \textbf{88}, 
015004 (2016).

\bibitem{Faria1}
F. F. Faria, Adv. High Energy Phys. \textbf{2014}, 520259 (2014).

\bibitem{Faria2}
F. F. Faria, Mod. Phys. Lett. A \textbf{36}, 2150115 (2021).

\bibitem{Faria3}
F. F. Faria, Adv. High Energy Phys. \textbf{2019}, 7013012 (2019).

\bibitem{Faria4}
F. F. Faria, Eur. Phys. J. C \textbf{80}, 645 (2020).

\bibitem{Faria5}
F. F. Faria, Mod. Phys. Lett. A \textbf{37}, 2250033 (2022).

\bibitem{Faria6}
F. F. Faria, Eur. Phys. J. C \textbf{76}, 188 (2016).

\bibitem{Faria7}
F. F. Faria, Eur. Phys. J. C \textbf{77}, 11 (2017).

\bibitem{Faria8}
F. F. Faria, Eur. Phys. J. C \textbf{78}, 277 (2018).

\bibitem{Faria9}
F. F. Faria, arXiv:1903.04893 [hep-th].

\bibitem{Matsuo}
N. Matsuo, Gen. Relativ. Gravit. \textbf{22}, 561 (1990).

\bibitem{Man1}
P. D. Mannheim, Gen. Relativ. Gravit. \textbf{22}, 289 (1990).

\bibitem{Hel}
A. J. Helmboldt, P. Humbert, M. Lindner and J. Smirnov, JHEP \textbf{2017},
113 (2017).

\bibitem{Man2}
P. D. Mannheim, Prog. Part. Nucl. Phys. \textbf{56}, 340 (2006).

\bibitem{Knox}
L. Knox and A. Kosowsky, arXiv:9311006 [astro-ph].

\bibitem{Tana}
M. Tanabashi et al. (Particle Data Group), Phys. Rev. D \textbf{98}, 030001 (2018).

\bibitem{Aver}
E. Aver et al., JCAP \textbf{03}, 027 (2021).

\bibitem{Zyla}
P. A. Zyla et al. (Particle Data Group), PTEP \textbf{2020}, 083C01 (2020).

\bibitem{Meissner}
K. A. Meissner and H. Nicolai, Phys. Lett. B \textbf{648}, 312 (2007).

\bibitem{Salv}
L. Salvati et al., JCAP \textbf{08}, 022 (2016).

\bibitem{Agui}
A. A. Aguilar-Arevalo et al. (MiniBooNE Collaboration), Phys. Rev. Lett. 
\textbf{121}, 221801 (2018).



\end{thebibliography}
\end{document}